# Oceanic Forcing on Interannual Variability of Sahel Heavy and Moderate Daily Rainfall

M. Diakhate[1*], B. Rodríguez-Fonseca[2,3], I. Gómara[4], E. Mohino[2], A. L. Dieng[1] and A. T. Gaye[1]

[1] Laboratoire de Physique de l'Atmosphère et de l'Océan – Siméon Fongang, Ecole Supérieure Polytechnique, Université Cheikh Anta Diop, BP 5085, 10700 Dakar, Senegal

[2] Departamento de Física de la Tierra y Astrofísica, Universidad Complutense de Madrid, Madrid 28040, Spain.

[3] Instituto de Geociencias (IGEO), UCM, CSIC, Madrid 28040, Spain

[4] CEIGRAM / Departamento de Producción Agraria, Universidad Politécnica de Madrid, Madrid 28040, Spain.

***Corresponding author:** Moussa Diakhate, Laboratoire de Physique de l'Atmosphère et de l'Océan – Siméon Fongang, Ecole Supérieure Polytechnique, Université Cheikh Anta Diop, BP 5085, 10700 Dakar, Senegal (moussa1.diakhate@ucad.edu.sn)

**Abstract:** This article analyzes SST remote forcing on the interannual variability of Sahel summer months (June to September) moderate (below $75^{th}$ percentile) and heavy (above $75^{th}$ percentile) daily precipitation events during the period 1981-2016. Evidence is given that interannual variability of these events are markedly different. The occurrence of moderate daily rainfall events appears to be enhanced by positive SST anomalies over the Tropical North Atlantic and Mediterranean, which act to increase low-level moisture advection towards the Sahel from equatorial and north tropical Atlantic (the opposite holds for negative SSTs anomalies). In contrast, heavy and extreme daily rainfall events seem to be linked to El Niño-Southern Oscillation (ENSO) and Mediterranean variability. Under La Niña conditions and a warmer Mediterranean, vertical atmospheric instability is increased over the Sahel and low-level moisture supply from the Equatorial Atlantic is enhanced over the area (the reverse is found for opposite sign SST anomalies). Further evidence suggests that interannual variability of Sahel rainfall is mainly dominated by the extreme events. These results have implications for seasonal forecasting of Sahel moderate and heavy precipitation events based on SST predictors, as significant predictability is found from 1 to 4 months in advance.

# 1 Introduction

The Sahel is facing an increase of extreme rainfall occurrence (Ly et al., 2013). Recently, Taylor et al. (2017) have shown that extreme thunderstorms over the Sahel are nowadays three times more frequent than they were 35 years ago. These extremes constitute the primary impact of climate change on society (Katz and Brown, 1992), more than changes in the mean climate (Mitchel et al., 1990). Due to the vulnerability of the Sahel region, precipitation extremes often directly (and indirectly) impact numerous economy sectors and are associated with elevated death tolls (New et al., 2006; Lobell et al., 2011; Anyamba et al., 2014; Sané et al., 2015; 2016). The predictability of these extreme events is a big challenge for the climate science community in the Sahel area.

Among processes controlling Sahel rainfall variability, oceanic forcing has been found to be the dominant driver (Giannini et al., 2003; Rowell, 2013; Rodríguez-Fonseca et al., 2011; 2015; Gómara et al., 2017). Many works have shown that seasonal rainfall variability over the Sahel is related to Sea Surface Temperature (SST) anomalies over the tropical Atlantic, Pacific and the Mediterranean Sea (Rodríguez-Fonseca et al., 2015; and references therein). In particular, a warming of Pacific and Atlantic tropical waters has been found to promote a decrease in seasonal Sahelian rainfall (e.g. Rowell, 2001; Janicot et al., 2001; Mohino et al., 2011a, b; Losada et al., 2010), while a warmer Mediterranean leads to an increase (Rowell, 2003; Fontaine et al., 2010). These relations are somewhat linear, as opposite responses in rainfall take place when colder than usual SSTs arise over the same marine regions.

Regarding Sahel daily rainfall intensity/frequency, Giannini et al. (2013) found that the recovery during the last two decades was accompanied by the relative predominance of an increase in median rainfall intensity. Such increase was shown to coincide with positive SST anomalies over the Tropical North Atlantic (TNA), potentially associated with a positive phase of the Atlantic Multidecadal Oscillation (AMO). Additionally, it was also shown that extreme dry spell occurrence over West Africa is significantly linked to anomalous SSTs in the Indian and south Atlantic oceans (Bader and Latif, 2011; Salack et al., 2013). Parhi et al. (2015) also revealed that El Niño-Southern Oscillation (ENSO) significantly influences the interannual variability of Sahel rainy days frequency during boreal summer.

In contrast, the possible influence of SST forcing on Sahel precipitation, stratified and isolated in terms of event intensity (e.g., moderate, extreme), has received little attention in the literature so far. Ta et al. (2016) recently endeavored this subject, but their analysis focused on tropical Atlantic SST forcing over the Sahelian and Guinean regions, and they looked for oceanic forcing associated with trends on extreme rainfall. Results from this study pointed out the existence of a significant link between extreme rainfall trends over West Africa and SST anomalies in the equatorial Atlantic.

Distinctively, the present paper focuses on the interannual variability of Sahel rainfall events of different intensities (e.g., low, moderate, heavy and extreme), searching for oceanic sources of predictability worldwide to provide skillful seasonal forecasts. The article is structured as follows. Section 2 describes the data and methodology used. Section 3 is centered in the analysis of Sahel moderate and heavy precipitation indices, their associated teleconnections and mechanisms and their seasonal predictability. A summary of the main findings concludes this article.

## 2. Data and Methodology

A daily rainfall dataset, the Climate Hazards Group Infrared Precipitation with Station data (CHIRPS), is utilized. The CHIRPS dataset spans from 1981 to present-day at 0.25°x0.25° longitude/latitude horizontal resolution and is based upon satellite infrared and rain gauge measurements over West Africa (Funk et al., 2014; 2015). Daily June to September (JJAS) precipitation measurements are considered to characterize Sahel precipitation indices over the area (20°W-10°E; 10°N-20°N). Note that similar results are obtained for daily July to September (JAS) measurements. Depending on percentile thresholds, indices are stratified in terms of daily rainfall intensity: Low (< $50^{th}$), Moderate (< $75^{th}$), Heavy (> $75^{th}$), and Extreme (> $95^{th}$) events. Therefore, the moderate category also includes low precipitation events. Likewise, the heavy category contains extreme daily events. Thresholds are calculated over each Sahel grid point, separately with rainy days defined as days in which the rainfall amount is greater than 1 mm. Subsequently, for each precipitation category, the grid-point precipitation values fulfilling the intensity criterion are retained and averaged over the total Sahel area for each JJAS season. Then, the seasonal anomalies of these indices are calculated by removing the seasonal mean. The obtained timeseries are standardized and high-pass filtered (Lanczos; 8-year cut-off period) to keep only with interannual variability (1 data point per year). The resultant timeseries are considered as our rainfall indices along the whole study.

The SST dataset from the Met Office Hadley Centre SST and sea ice data (HadISST) is used and regressed on precipitation indices to search for possible oceanic teleconnections. The SST data has a horizontal grid of 1°x1° in longitude/latitude (Rayner et al., 2003). The indices of El Niño–Southern Oscillation (ENSO), TNA and Mediterranean Sea Index (MSI) are computed based on high-pass filtered SST anomalies, averaged over the regions (180°W-80°W, 5°S-5°N), (60°W-20°W, 5°N-30°N) and (5°W-30°E, 30°N-45°N), respectively.

To analyze the underlying physics of the identified teleconnection mechanisms, atmospheric data from the European Center for Medium-Range Weather Forecasts (ECMWF) ERA-Interim reanalysis (period 1979-2017) is also utilized. ERA-Interim is provided at 0.75°x0.75° horizontal resolution with 60 vertical atmospheric levels, from surface up to 0.1 hPa (Dee et al., 2011). The dynamical fields analyzed to compute JJAS seasonal anomalies are: (i) specific humidity at 850 hPa; (ii) moisture flux at 850 hPa; and (iii) vertical instability: difference of wind divergence between 200 and 850 hPa pressure levels (hereafter DIV200/850; Gómara et al., 2017). Positive DIV200/850 values thus indicate vertical destabilization and upward motion of air (the reverse is true for negative values). The Sea Level Pressure (SLP) is also used (instead of the DIV200/850 parameter) for the MSI, as the influence of the Mediterranean Sea in Sahelian rainfall has been found to take place through anomalous moisture advection at surface levels, mediated by the SLP gradient between the Gulf of Guinea and Sahara, as shown in previous studies (Rowell, 2003; Fontaine et al., 2010; Gómara et al., 2017).

To identify co-variability between fields, regression maps are calculated by projecting a time evolving field onto the timeseries associated with a particular index. For each grid point (i,j), a linear regression is done between an index I, which is standardized (for example rainfall or TNA), and the value of anomalous of the variable Y at that grid point (1):

$$R(i,j) = \sum_{t=1}^{n} Y(i,j,t) I(t) \qquad (1)$$

To further analyze the atmospheric stability associated with moderate-weak and heavy-extreme precipitation events over the Sahel, Moist Static Energy (MSE) is calculated (2):

$$\text{MSE} = \underbrace{C_P T}_{SH} + \underbrace{gz}_{PE} + \underbrace{L_v q}_{LH} \qquad (2)$$

where T is the temperature; Cp, the specific heat at constant pressure; z, the height; g, the gravitational acceleration; $L_v$, the latent heat of vaporization at 0°C; and q, the specific humidity. MSE thus combines the contributions from latent heat (LH), sensible heat (SH) and potential energy (PE) terms. The MSE is a useful tool to investigate the relative roles of moisture and temperature changes to this stability (Sheen et al. 2017; Hill et al. 2017). An increasing MSE with altitude denotes a stable atmosphere.

Due to the time limitations posed by the observational datasets, the time interval considered for the study is 1981-2016, a period in which a strong Sahel-ENSO teleconnection has been documented (Janicot et al., 2001; Mohino et al., 2011a; Rodriguez-Fonseca et al.; 2011; 2015; Gómara et al. 2017). To analyze the predictability of Sahel daily precipitation events based on SST anomalies, a stepwise multi-linear regression statistical model is utilized for the same period (cf. Section 3.4). Finally, a two-tailed t-test that accounts for auto-correlation of the timeseries (Bretherton et al., 1999) is used throughout the paper for testing statistical significance of the results (95% confidence interval).

## 3 Results

### 3.1 Interannual variability of moderate and heavy daily Sahel rainfall

Total Sahelian JJAS rainfall variability is separated into the contributions of low, moderate, heavy and extreme daily rainfall events. A first result arises when performing this separation, as moderate and heavy daily rainfall variability indices show very different interannual time evolutions, appearing to be caused by distinctive physical mechanisms. Figure 1 shows that, except for some periods and years like 1993-1996, 1998-2002, 2014 and 2015, the time evolution of these indices is decoupled. This is especially evident before 1993 and after 2005.

Correlation values between low, moderate, heavy and extreme indices are presented in Table 1. Quantitatively, correlation values between moderate and heavy indices are found to be positive (0.24), but not statistically significant, supporting therefore the conclusion that variability of moderate and heavy rainfall events over the Sahel is decoupled.

To quantify the imprint of each of the rainfall intensity categories on the total rainfall index, their correlation values are provided in Table 2, together with their standard deviations. The correlation value between moderate and total precipitation indices is 0.46, and that between heavy and total precipitation is 0.97. Consequently, interannual variability of total Sahel rainfall is found to be mainly dominated by heavy daily precipitation events. However, regarding the contribution to yearly precipitation amount, both heavy and moderate rainfall are important (51% and 49% of the total annual precipitation, respectively). For the remaining categories, the contribution of low daily rainfall events to the total precipitation is found to be more important in terms of yearly rainfall accumulation (23%) than in terms of explained interannual variability (cf. Table 2; right column).

**3.2 SST teleconnection patterns**

Consistent with previous studies (e.g. Rowell, 2001; Janicot et al., 2001; Mohino et al., 2011b; Fontaine et al., 2010; Rodriguez-Fonseca et al., 2011), an increase of JJAS Sahel total rainfall is associated with a La Niña-like pattern in the equatorial Pacific, and warmer than usual SSTs over the Mediterranean and subtropical North Atlantic (Figure 2a). By linear construction, the opposite is also true for a decrease in total seasonal rainfall. Figures 2b, c, d respectively show JJAS SST forcing patterns associated with moderate, heavy and extreme rainfall events. In this context, please note that similar SST patterns have been found considering slight variations in the total study period (1981-2011, 1981-2012 or 1981-2013). The enhanced occurrence of moderate daily Sahelian rainfall events is related to a SST warming of the tropical/subtropical North Atlantic (Figure 2b), whilst a La Niña-like signal appears over the equatorial Pacific associated with heavy and extreme daily rainfall events (Figures 2c-d). Whereas correlation values between TNA/moderate (0.67) and ENSO/extreme (-0.35) rainfall events are statistically significant (95% confidence interval), TNA/extreme (-0.04) and ENSO/moderate (0.15) not (Table 3). Over the North Atlantic a horseshoe pattern, typically associated with the negative phase of the North Atlantic Oscillation (NAO), is also observed in Figure 2b. Under a negative NAO phase, the weakening of the trade winds decreases the latent heat flux and sea level pressure over subtropical latitudes, warming the oceanic waters beneath through the WES (wind-evaporation-SST) feedback process (Okumura et al., 2001; Czaja and Frankignoul, 1999). The Mediterranean (MSI) also shows a significant link with both moderate and heavy/extreme events (correlation values of 0.67, 0.62 and 0.31 for moderate, heavy and extreme events; cf. Table 3). The patterns associated with heavy (Figure 2c) and extreme daily precipitation events (Figure 2d) mostly resemble that associated with total rainfall (Figure 2a). Nevertheless, the intensity/extent of the SST anomalies over the Pacific appears enhanced for the case of heavy/extreme daily precipitation events, pointing out to a tropical Pacific forcing as a main driver of intense rainfall, compared to the secondary role of extratropical patterns (North Atlantic and Mediterranean; Figures 2c, d).

**3.3 Mechanisms of SST teleconnections on Sahel moderate and heavy daily rainfall events**

In Figure 3 the underlying physical mechanisms of the teleconnections identified in Figure 2 are analyzed. Figure 3a shows that Sahel total rainfall variability is associated with anomalous southwesterly low-level moisture flux, which transports humidity from the north tropical and equatorial Atlantic towards the Sahel (arrows). The moisture advection enhances the available low-level humidity over this area (shadings). Positive vertical instability (DIV200/850) anomalies are also observed over the same region for total rainfall, reflecting low-level wind convergence and upper level divergence associated with vertical air movement and, potentially, convection (contours).

A similar picture is observed for heavy and extreme daily precipitation events (Figures 3c, d). Organized deep convective systems are characterized by a baroclinic structure with opposite divergent circulation patterns at upper levels, a feature denoted in contours. The maximum anomalies of low-level humidity are located over the central and eastern Sahel (Figures 3c-d). This flow is driven by the large-scale circulation, and probably associated with La Niña, which acts to promote vertical air destabilization over the Sahel (Joly and Voldoire, 2009).

Conversely, for moderate daily rainfall events (Figure 3b), local negative DIV200/850 anomalies are present over the Sahel region, suggesting that local convection is not enhanced for these events. There is, however, enhanced low-level humidity stemming from the TNA region, which is consistent with a warming in the tropical/subtropical North Atlantic waters (Figure 2b). Such warming acts to increase evaporation over these areas and therefore moisture advection towards the Sahel.

These results indicate that heavy and extreme daily rainfall events seem to be mainly associated with ENSO, whilst moderate events seem to be linked to SSTs anomalies from the TNA. SST variability over the Mediterranean appears to be important for both moderate and heavy-extreme rainfall (cf. Figure 2b-d). However, regarding the mechanism, Figures 3b-c show no clear significant low-level humidity flux from the Mediterranean towards the Sahel.

To test the robustness of the main dynamical mechanisms proposed, a reverse methodological approach is carried out in Figure 4. First, worldwide SST anomalies re regressed onto ENSO, TNA and MSI indices (Figures 4a, b, c). Second, atmospheric variables from Figure 3 are also regressed on TNA and ENSO indices (Figures 4d, e). The same analysis is done for the MSI index by just replacing the DIV200/850 parameter by SLP, as explained before (Figure 4f).

For a TNA positive phase (as for moderate rainfall events), the regression map shows enhanced low-level specific humidity over the Sahel region, transported from the tropical/subtropical North Atlantic (Figure 4d). In this area, a horseshoe pattern (Figure 4a), together with a negative phase of the NAO, may act to reduce the trade winds, thus increasing humidity advection over the Sahel (from a warming of the TNA).

For an ENSO negative phase (as for Sahel heavy/extreme events), results show enhanced north-eastward low-level moisture flux, transporting humidity from the northern tropical and Equatorial Atlantic towards the Sahel. Such anomalous flow is again consistent with the vertical destabilization (stabilization) that La Niña (El Niño) induces over the Sahel, promoting wind convergence (divergence) at 850 hPa and divergence (convergence) at 200 hPa (contours in Figure 4e). Figure 4b also reveals that an equatorial Pacific cooling (warming) is associated with a concomitant equatorial Atlantic warming (cooling). The simultaneous appearance of these anomalies for the period considered (1981-2016) has been reported in previous studies (Rodríguez-Fonseca et al. 2009).

Finally, for MSI, Figure 4f shows that warmer Mediterranean enhances evaporation and strengthens the Sahara heat low, thus increasing the pressure gradient between the Sahara and Gulf of Guinea in agreement to previous studies (Rowell, 2003; Peyrillé et al., 2007; Fontaine et al., 2010; Gaetani et al. 2010; Gómara et al., 2017). The increased meridional pressure gradient promotes a further penetration of the monsoonal flow from the Gulf of Guinea (anomalous southwesterly positive moisture flux). Figure 4f also shows that this meridional circulation strengthens with changes in the heat-low, combined with a large change of westerly flow from the tropical North Atlantic, that also increases moisture and heat transport into the Sahel (Martin and Thorncroft, 2014). In addition, and as it will be shown later, the inclusion of the MSI leads to enhanced predictive skill (cf. Section 3.4).

To further analyze the origin of rainfall variability and its relation to intensity, an analysis of moist static energy (MSE) (2) is subsequently performed in Figure 5. For the variability of moderate and extreme daily precipitation events, the contribution of the Potential Energy component (PE; Figure 5d) is negligible compared to the Latent Heat (LH) and Sensible Heat (SH) terms (Figures 5b and 5c). Attending to moderate and extreme daily precipitation vertical profiles, the stronger differences arise in the low tropospheric levels (Figures 5a, b).

MSE associated with low-moderate daily precipitation decreases strongly with altitude compared to that associated with extremes (Figure 5a), which suggests instability in the low levels (Sheen et al., 2017). Figure 5c shows that this strong decrease is mainly explained by the LH component. However, the LH component decreases with a similar rate for extremes too. Surface evaporation is important over land where soil moisture is available, a feature that is very important in moderate rainfall events, where MSE is controlled by LH. In contrast, during extreme daily precipitation events, the contribution of LH component (Figure 5c) is balanced at low tropospheric levels by strong negative values of SH (Figure 5b). These strong negative SH values could depict the equilibrium response once convection has overturned the vertically-integrated temperature column (Wei et al., 2014). At medium and upper levels (500-200 hPa), the MSE profile is more unstable for extreme than for moderate precipitation events (Figure 5a). This can be related to both, LH and SH components. On the one hand, the upper troposphere (400 to 200 hPa) is colder in the extreme case, consistent with a tropical cooling due to La Niña (Lintner and Chiang, 2007), which affects the SH component. On the other hand, the middle troposphere is more humid in the extreme case, which would be related to a stronger vertical rise of moist rich air (Figure 3d).

In summary, moderate events seem to be controlled by instability in lower levels associated with changes in latent heat fluxes due to surface evaporation. The influence of the subtropical North Atlantic and Mediterranean SSTs is felt through surface processes and moisture flux advection from the warm seas. In contrast, extreme rainfall events are controlled by upper level instability and convective precipitation systems, a mechanism consistent with ENSO remote influence, as the teleconnection with ENSO is related to changes in the upper level divergence, favoring (La Niña) or inhibiting (El Niño) the instability in the upper levels.

This hypothesis is confirmed in Figure 6, where air temperature anomalies from surface up to 200 hPa are regressed on moderate and extreme precipitation indices and averaged over the corridor 0°N-17°N. Similar result (but with a weaker signal) are obtained when limiting the latitudinal band to Sahelian one (10-20ºN). The extended band has been therefore chosen in order capture a better ENSO signature, as the teleconnection associated with the latter is stronger near the equator.

During moderate daily rainfall events, as shown by the SH anomalies profile (Figure 5b), atmospheric temperature anomalies are significantly warmer over the tropical north Atlantic (Figure 6a). These results are consistent with the presence of warmer waters over the near subtropical north Atlantic, which heat up the air immediately above. During extreme daily rainfall events, in addition to weak upper level cooling over the Sahel, associated with La Niña influence (the opposite takes place for El Niño), the temperature anomalies exhibit a strong and significant low-level cooling in the lower levels (Figure 6b). This cooling could be associated with the temperature equilibrium response, once convection has overturned the column (Figure 5b). Consequently, MSE is quasi-constant with altitude for extreme daily precipitation at lower levels, due to the balance of SH and LH terms. In this context, it is well known that the evaporation/sublimation of rain/ice falling droplets produces a strong cooling in convective systems, which subsequently strengthens the downward current itself and cold pools. These latter are considered as fundamental ingredients for deep convection (Torri et al., 2015). However, due to their mesoscale characteristics, their direct causality on seasonal and spatial averaged rainfall indices from this study is extremely hard to detect (Wei et al. 2014).

The results from this section are entirely consistent with the same analysis applied to non-filtered data (linear trends removed), as the SST patterns considered are mainly interannual (ENSO, TNA, Mediterranean).

**3.4 Predictability of Sahel moderate and heavy/extreme rainfall at inter-annual timescales**

As described in the previous section, the key SST indices associated with JJAS Sahel rainfall variability at inter-annual timescales are the Tropical North Atlantic (TNA), the El Niño Southern Oscillation (ENSO) and the Mediterranean Sea (MSI).

To assess the potential predictability of rainfall events over the Sahel, different predictors are selected based on these SST indices. Potential predictors for each Sahel rainfall index are identified by applying a stepwise multi-linear regression. This method permits to automatically identify the relevant SST predictors for each rainfall index and disregard those which do not increase model skill (forward selection and backward elimination, von Storch and Zwiers 1999). Subsequently, the selected predictors for each index are utilized to perform a leave-one-out cross-validated hindcast at lag 0 (predictors/predictands based on the same monthly period: August). In this way, the linear regression model is built for each year to be predicted, calculating the coefficients of the model with all the years in our database except the one that is predicted (ter Braak and Juggins, 1993; Birks, 1995). The hindcast is performed and correlated with the omitted observations. The same analysis is extended by utilizing the same SST predictors from the previous months to assess predictability of Sahel moderate and heavy daily rainfall events. Unlike in the previous analyses, all considered indices and anomalies in this section are calculated on a monthly basis. As a result, all selected rainfall predictands (precipitation indices) are computed for the month of August. SST predictors considered for lag 0 are thus simultaneous SST anomalies. Lag-1 forecasts take into account SST anomalies from the previous month of the same year (July), and so on back to one year in advance.

In Table 3 the main SST predictors utilized to predict rainfall indices at lag 0 are provided. As expected, the regression model considers the TNA as main predictor of moderate Sahel rainfall, together with the MSI. For heavy/extreme precipitation, both ENSO and MSI are considered as skillful predictors. This outcome is fully consistent with the findings from Figures 2 to 6.

Based on these predictors, the cross-validation hindcast for each rainfall index at lag 0 is provided (Figure 7). The model is able to reproduce the temporal evolution of Sahel rainfall indices with high significant skill (0.73/0.69/0.50 anomaly correlation coefficients for moderate/heavy/extreme rainfall indices; 95% confidence interval).

In Figure 8 the same analysis is extended to consider SST predictors from lag 0 (August) to -12 (August/yr-1). In the figure, lagged correlations between rainfall predictands and SST predictors are provided with bars, whereas skill scores from the multi-regression model are provided with markers. For moderate rainfall events, significant forecast skill is obtained from MSI and TNA predictors 2 months ahead (Fig. 8 top; 95% confidence level). In other words, the model is able to predict August moderate rainfall based on MSI and TNA SST predictors from June.

This outcome is also consistent with lagged correlation values, which are statistically significant (same confidence interval) for both MSI (thin bars) and TNA (wide bars) back to lag -2. For heavy and extreme rainfall indices the forecast window (based on MSI and ENSO SST predictors) is even increased to 3 and 4 months in advance (May/April), respectively (Figures 8 central and bottom). Also, the lagged correlations with the MSI index decrease more rapidly backwards in time than for ENSO. Correlation values between ENSO and the rainfall indices remain significant from April on (lag -4), whereas those of MSI just reach the previous July (lag -1). The fact that the Mediterranean variability is stronger during the summer months may be behind this behavior (Gaetani et al. 2010). Finally, no correlation is found between August heavy/extreme precipitation anomalies and ENSO during the previous winter (December/January), suggesting that the observed teleconnection takes place during the onset phase of La Niña/El Niño.

## 4 Conclusions and discussion

The contributions, variability and teleconnections of different types of precipitation, stratified in terms of intensity, have been analyzed over the Sahel. For this purpose, a precipitation dataset (CHIRPS), a sea surface temperature database (HadISST) and a reanalysis product (ERA-Interim) have been analyzed during the summer boreal season (June to September) along 36 consecutive years (1981-2016). Rainfall categories have been built upon grid-point percentile thresholds during rainy days (days with precipitation amount greater than 1 mm) as follows: Low ($< 50^{th}$), Moderate ($< 75^{th}$), Heavy ($> 75^{th}$), and Extreme ($> 95^{th}$) daily rainfall events. For each category, precipitation values have been retained and averaged over the area (20ºW-10ºE; 10ºN-20ºN). Lastly, seasonal precipitation anomalies have been calculated and a high-pass filter has been applied to the timeseries to keep only the interannual variability (allowed oscillation periods from 1 to 8 years).

Evidence has been presented that total Sahelian rainfall variability is dominated by the occurrence/absence of heavy and extreme daily rainfall events, rather than moderate or low rainfall episodes, whose variability and teleconnections are markedly different (Tables 1-2 and Figure 1).

On the one hand, Sahel heavy and extreme precipitation variability is mainly linked to El Niño-Southern Oscillation (ENSO) and, to a lower extent, Mediterranean SST anomalies (MSI). During La Niña episodes (Figures 2c,d), the induced anomalous circulation over the Sahel enhances vertical destabilization, promoting air convergence at lower levels and divergence at upper levels (Figures 3c,d and 5). Such conditions, together with the penetration of moist rich air from the Gulf of Guinea towards the Sahel (Figures 3c,d), create a very favorable environment for convection and thus heavy and extreme rainfall occurrence (Figures 5 and 6b). The reverse holds during El Niño and negative MSI episodes.

On the other hand, moderate rainfall variability is primarily associated with tropical North Atlantic (TNA) and also Mediterranean SST anomalies (Figure 2b). When these SST regions are warmer than average, enhanced moisture transport from the equatorial and north tropical Atlantic towards the Sahel is triggered (Sheen et al., 2017), producing an increase of humidity availability over the region (Figure 3b). In this case, the vertical atmospheric stability is reduced over the Sahel (Figure 3b) and environmental conditions may be favorable for the formation of horizontal stratiform clouds (Figures 5 and 6a), potentially responsible of the moderate precipitation events. As before, these conditions are reversed for negative TNA and Mediterranean SST anomalies.

The Atlantic influence is associated with enhanced available humidity in the lower and medium levels of the troposphere. The main difference with the Pacific forcing is not the moisture advection, but the vertical destabilization of the atmosphere. For the latter, air uplift is promoted and, therefore, so are heavy precipitation events from deep convection, including organized mesoscale or synoptic systems such as Mesoscale Convective Systems or African Easterly Waves (AEWs). According to Okonkwo (2014), la Niña event over the Pacific increases the frequency of AEWs. One can therefore expect a stronger link between extreme rainfall events and AEWs. These AEWs are a major source of synoptic-scale rainfall variability throughout West Africa (Skinner and Diffenbaugh 2013, Martin and Thorncroft 2015). Following the approach from Diedhiou et al. (1998) for the identification of AEWs, in our case the number of AEWs associated with heavy rainfall (22%) is two times higher than that associated with moderate ones (11%; cf. Table 4).

In terms of yearly precipitation amounts, moderate and heavy daily rainfall contributions are of the same order of magnitude (ca. 50%; Table 2). This is an important result when tackling, for instance, climate services and impacts in agriculture, as heavy daily precipitation events have more impact than extreme ones and contribute more on interannual variability (cf. Table 2 – first column). Some signal over the Mediterranean has been found in relation to heavy daily rainfall events (Figure 2c), thus suggesting that these results might complement those relating recent trends in Mediterranean warming on Sahelian recovery (Park et al., 2016; Biasutti, 2016).

Finally, a stepwise multi-linear regression model has been applied to identify the most relevant SST predictors of each rainfall index at different forecast windows. For this purpose, indices of moderate, heavy and extreme daily precipitation have been computed based on August monthly anomalies (predictands). Monthly SST anomaly indices (ENSO, TNA and MSI) from the same month (lag 0), and the previous ones (back to lag -12), have been considered as predictors. Results indicate that moderate, heavy and extreme Sahel rainfall anomalies can be predicted with high confidence utilizing SST predictors several months in advance. TNA and MSI predictors from two months in advance, together, are able to provide model skill for the prediction of moderate August Sahelian rainfall. Furthermore, Sahel heavy and extreme precipitation anomalies can be predicted up to 3 and 4 months ahead, respectively, considering as predictors ENSO and MSI. The findings provided in this article may have implications on seasonal forecasting of Sahel moderate, heavy and extreme rainfall daily events.

In this sense, it must be noted that due to the relative short time period considered (36-yr), SST-rainfall links obtained from this study are assumed to remain stable in time. To analyze the impact of non-stationarity in these relationships (Janicot et al., 2001; Mohino et al., 2011a; Rodriguez-Fonseca et al., 2011; 2015; Losada et al., 2012; Suárez-Moreno et al., 2018), a longer period rainfall dataset should be required. Finally, an analogous analysis based on regional model data (e.g., CORDEX-Africa), which considers potential non-linearities in the SST-precipitation relationships, will be another interesting path for future research.

**Acknowledgments**

The research leading to these results has received funding from: the CSIC ICOOP project, ICOOPB20204, the UCM Cooperation Project #7 from XIV Call; the NERC/DFID Future Climate for Africa programme under the AMMA-2050 project, grant number NE/M020428/1; the Spanish Project CGL2017-86415-R; and the EU/FP7 PREFACE project under grant agreement no. 603521. Iñigo Gómara is supported by the Spanish Ministry of Science and Competitiveness (Juan de la Cierva-Formación contract FJCI-2015-23874) and Universidad

Politécnica de Madrid (Programa Propio - Retención de Talento Doctor). The Climate Hazards Group InfraRed Precipitation with Station data (CHIRPS) is available from the Climate Hazards Group (http://chg.geog.ucsb.edu/data/chirps/). The Hadley Centre Sea Ice and Sea Surface Temperature data set (HadISST) is available from the UK Met Office (https://www.metoffice.gov.uk/hadobs/hadisst/). The European Center for Medium-Range Weather Forecasts (ECMWF) ERA-Interim reanalysis (period 1979-2017) is available from the ECMWF MARS server (https://www.ecmwf.int/en/research/climate-reanalysis/era-interim). We would like to acknowledge Julián Villamayor and Chris Taylor for constructive remarks. Finally, we would like to thank the three anonymous reviewers for their helpful comments and suggestions, which have contributed to improve this manuscript.

**Table 1**: Cross-correlations of Sahel daily rainfall (low, moderate, heavy and extreme) indices. Boldface indicates statistical significance at the 95% confidence interval using a Student t test.

| Correlation | Moderate | Heavy | Extreme |
|---|---|---|---|
| **Low** | **0.64** | -0.19 | -0.12 |
| **Moderate** |  | 0.24 | -0.04 |
| **Heavy** |  |  | **0.89** |

**Table 2**: Correlation values between the total rainfall index and that of low, moderate, heavy and extreme rainfall provided in the first column. Standard deviation coefficients of each index provided in column 2 and the contribution to yearly precipitation amount of each of these categories in column 3 (calculated by divided total rainfall accumulation associated with each category by the that associated with the total one). Boldface indicates statistical significance at the 95% confidence interval using a Student t test.

|  | Correlation with the total index | Std (mm) | % of contribution to yearly amounts |
|---|---|---|---|
| Total | **1.00** | 0.28 | 100 |
| Low (<50$^{th}$) | -0.02 | 0.03 | 23 |
| Moderate (<75$^{th}$) | **0.46** | 0.06 | 49 |
| Heavy (>75$^{th}$) | **0.97** | 0.25 | 51 |
| Extreme (>95$^{th}$) | **0.80** | 0.14 | 16 |

**Table 3**: Stepwise regression of the 3 main predictors (MSI, TNA and Nino3) onto Sahel August moderate, heavy and extreme rainfall indices at lag 0. Stepwise regression is the iterative application of forward selection and background elimination; it allows us to select best predictor for each Sahel rainfall index. Best predictors are associated with the status "in", and those that do not increase the model skill are associated with the status "out". The correlation values between SST indices and that of Sahel rainfall are also added in parentheses (significant threshold of 0.28 according to the Student test at 95%).

|  | MSI | TNA | Nino3 |
| --- | --- | --- | --- |
| Moderate (<75pctl) | In (0.67) | In (0.67) | Out (0.04) |
| Heavy (>75pctl) | In (0.62) | Out (0.44) | In (-0.30) |
| Extreme (>75pctl) | In (0.31) | Out (0.15) | In (-0.35) |

**Table 4:** African Easterly Waves (AEWs) identification. An AEWs is detected when the 2-9-days pre-filtered meridional winds at 700 hPa are greater than the value of its standard deviation at 0ºW-17.5ºN. Intersection at lag 0 with moderate, heavy and extreme daily rainfall occurrence and the passage of AEWs have been applied to estimate values within the table.

| Period 1981-2016 | Moderate | Heavy | Extreme |
| --- | --- | --- | --- |
| Total number of daily events | 11817 | 1240 | 248 |
| Total number of daily events coinciding with AEWs | 1403 | 278 | 40 |
| % | 11.87 | 22.42 | 16.13 |

**Figure Captions**

**Figure 1**: Normalized Sahel JJAS total (black), low (<50th percentile, red) moderate (<75th percentile, green), heavy (>75th percentile, blue) and extreme (>95th percentile, blue dashed and squared line) daily rainfall indices. These indices are built from the CHIRPS datasets and high-pass filtered. For each precipitation category (associated with a percentile), the grid-point precipitation values fulfilling the intensity criterion are retained and averaged over the total Sahel area (20°W:10°E/10°N:20°N) for each JJAS season.

**Figure 2:** Regression map of JJAS (1981-2016) SST anomalies (in °C per standard deviation of the rainfall index) onto total (a), moderate (b), heavy (c) and extreme (d) normalized precipitation indices. Contours delimit the 95% confidence interval using a Student t test. The black box over West Africa delimits the Sahel region considered in this paper.

**Figure 3**: Regression of JJAS (1981-2016) anomalies of specific humidity at 850 hPa (colors, in kg/kg), DIV200/850 (contours, $10^{-7}$ s$^{-1}$) and moisture flux at 850 hPa (arrows, kg/kg m/s) onto total (a), moderate (b), heavy (c) and extreme (d) normalized precipitation indices. Heavy contours delimit the 95% confidence interval using a Student t test. The black box over West Africa delimits the Sahel region considered in this paper.

**Figure 4:** Global JJAS SST (a,b,c) anomalies regressed onto TNA (a), ENSO (b) and MSI (c) indices (in °C per standard deviation index). Panels (d) and (e) present anomalies of specific humidity at 850 hPa (colors, in kg/kg), DIV200/850 (contours, 0.3 10-6 s-1 CI) and moisture

flux at 850 hPa (black and grey arrows for significant and non-significant anomalies respectively, kg/kg m/s) regressed onto TNA and ENSO indices respectively. Panel (f) corresponds to anomalies of specific humidity at 850 hPa (colors, in kg/kg), sea level pressure (contours, 3 Pa CI) and moisture flux at 850 hPa (black and grey arrows for significant and non-significant anomalies respectively, kg/kg m/s) regressed onto MSI index. Note that for panels (b) and (e), ENSO index is here multiplied by -1 in order to catch the la Niña effect. Contours in plots a and b delimit the 95% confidence interval using a Student t test, while in plots c and d only significant anomalies have been plotted. The continental box in (d), (e) and (f) delimits the Sahel region while boxes in panels (a), (b) and (c) correspond to TNA, ENSO and MSI indices areas respectively.

**Figure 5**: Regression of JJAS (1981-2016) anomalies of a) moist static energy (MSE) and its components: b) sensible heat (SH), c) latent heat (LH) and d) potential Energy (PE) onto moderate (line) and extreme (circles) normalized rainfall indices. The regressions onto the Sahel indices are done in each grid point of the Sahel box, and the vertical profiles correspond to anomalies averaged over the Sahel box.

**Figure 6**: Regression of JJAS (1981-2016) vertical (1000-200 hPa) temperature anomalies onto moderate (a) and extreme (b) normalized rainfall indices. Temperature anomalies are averaged over the latitudinal corridor 0°-17°N for each pressure level (colors, in °C). Black dotted contours delimit significant anomalies at 95% confidence interval. Vertical black dotted bars delimit longitudinal boundaries of the Sahel region.

**Figure 7**: Cross-validated hindcast of Sahel August moderate (top), heavy (middle) and extreme (bottom) rainfall indices based on predictors selected from the stepwise regression (table 4). Note that the significant threshold for the model skill is equal to 0.2752.

**Figure 8:** Lagged correlation and cross-validation hindcast of Sahel moderate (top), heavy (middle) and extreme (bottom) rainfall indices for August. Bars denote lagged correlations on monthly between August precipitation indices (predictands) and preceding SST anomaly indices (predictors) from Table 4. The dashed horizontal line delimits the statistical significance (95% confidence level) for correlation. Markers are plotted whenever the regression model skill is statistically significant: crosses (predictors TNA and MSI considered), stars (predictors Nino3*(-1) and MSI considered). Note that the legend is the for panels (b) and (c).

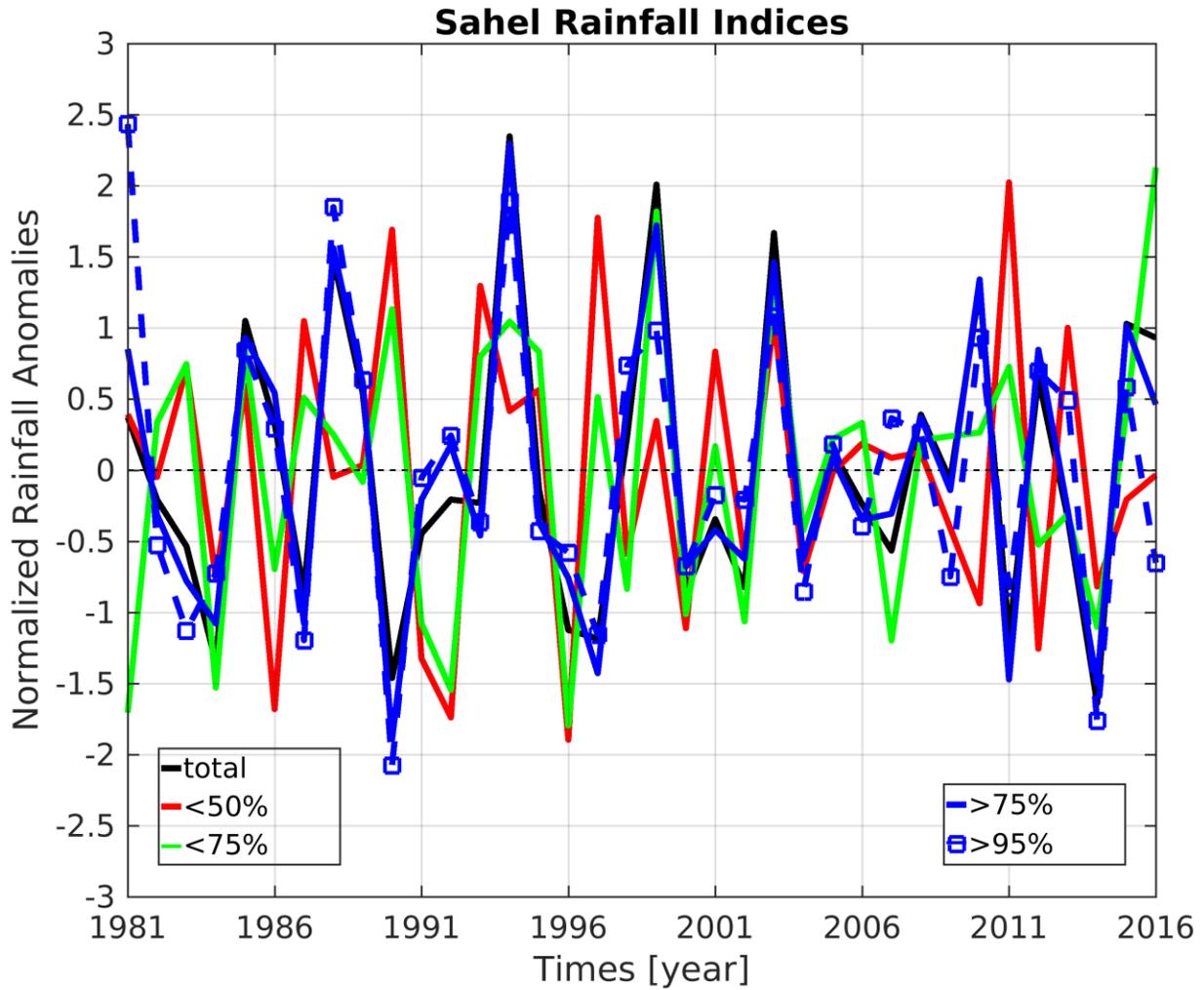

**Figure 1**: Normalized Sahel JJAS total (black), low (<50th percentile, red) moderate (<75th percentile, green), heavy (>75th percentile, blue) and extreme (>95th percentile, blue dashed and squared line) daily rainfall indices. These indices are built from the CHIRPS datasets and high-pass filtered. For each precipitation category (associated with a percentile), the grid-point precipitation values fulfilling the intensity criterion are retained and averaged over the total Sahel area (20ºW:10ºE/10ºN:20ºN) for each JJAS season.

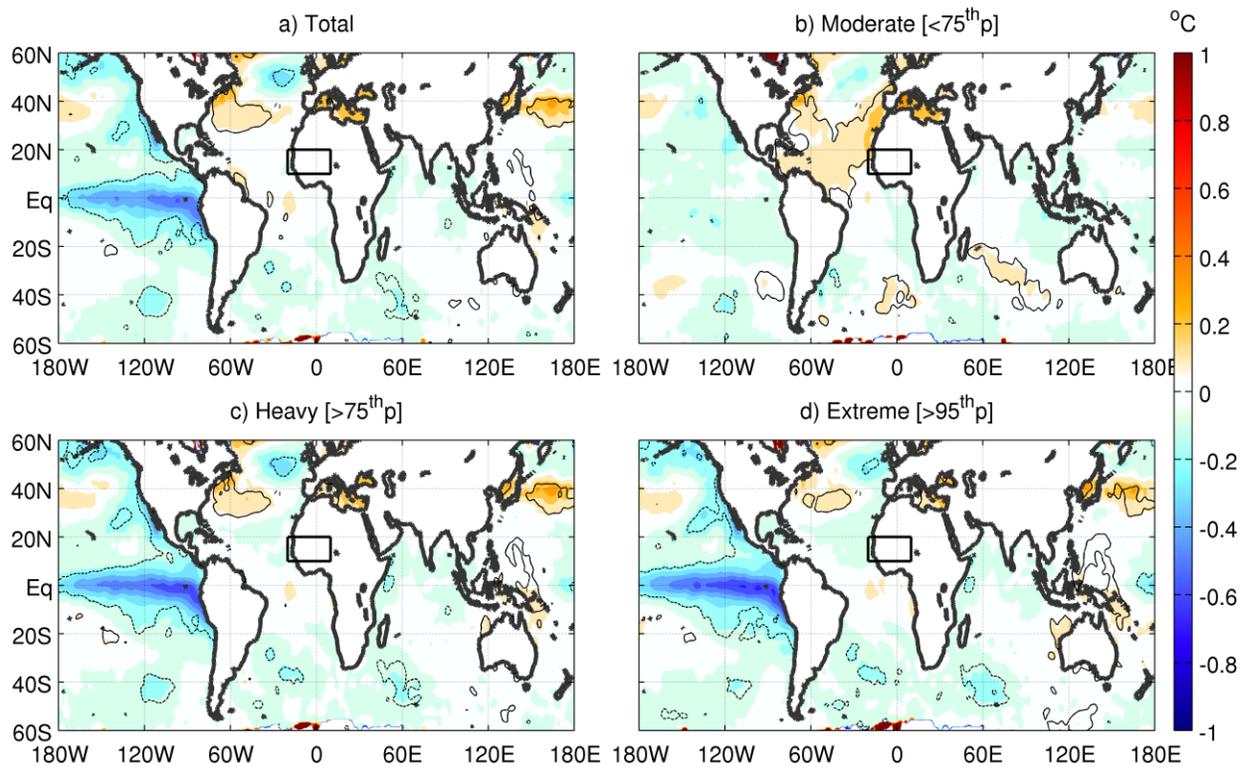

**Figure 2:** Regression map of JJAS (1981-2016) SST anomalies (in °C per standard deviation of the rainfall index) onto total (a), moderate (b), heavy (c) and extreme (d) normalized precipitation indices. Contours delimit the 95% confidence interval using a Student t test. The black box over West Africa delimits the Sahel region considered in this paper.

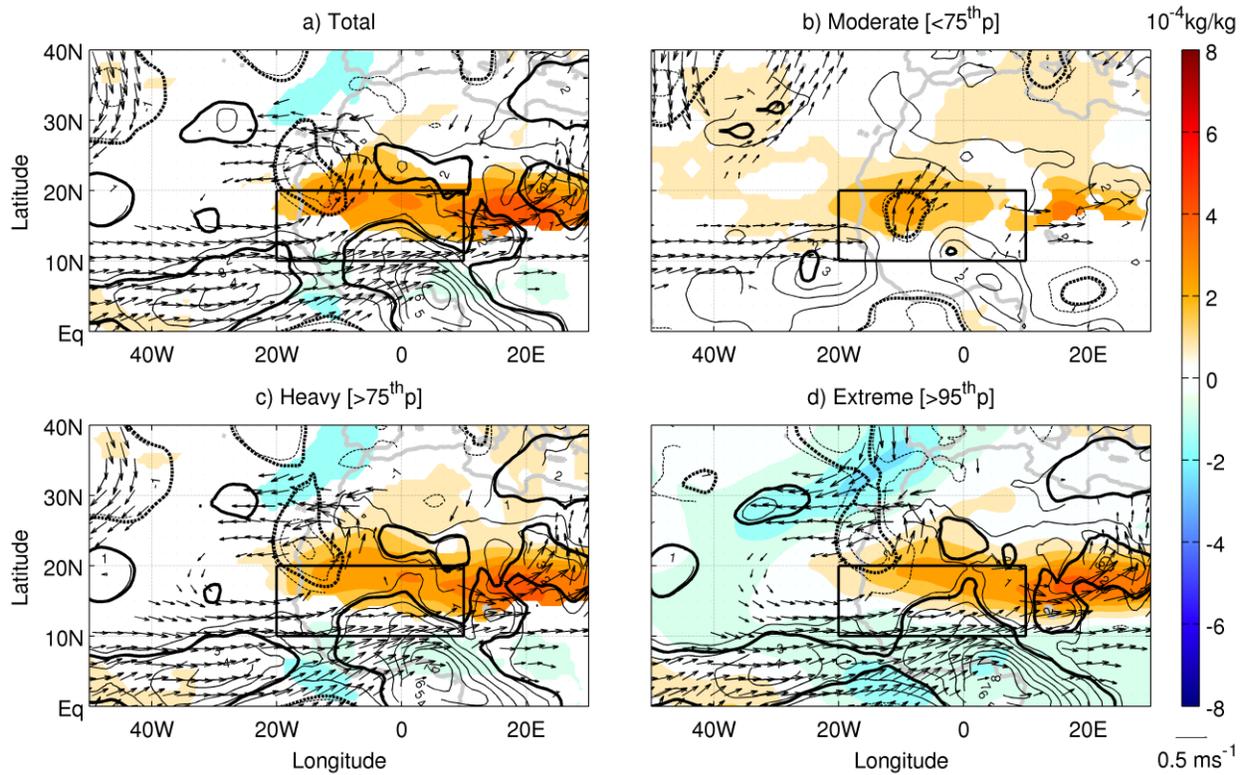

**Figure 3**: Regression of JJAS (1981-2016) anomalies of specific humidity at 850 hPa (colors, in kg/kg), DIV200/850 (contours, $10^{-7}$ s$^{-1}$) and moisture flux at 850 hPa (arrows, kg/kg m/s) onto total (a), moderate (b), heavy (c) and extreme (d) normalized precipitation indices. Heavy contours delimit the 95% confidence interval using a Student t test. The black box over West Africa delimits the Sahel region considered in this paper.

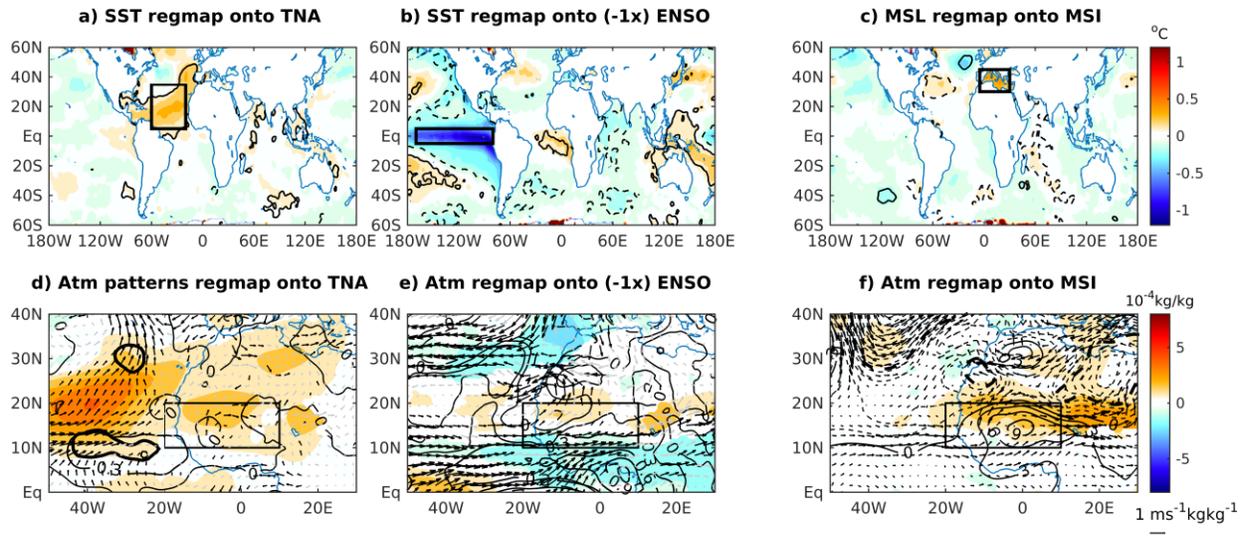

**Figure 4:** Global JJAS SST (a,b,c) anomalies regressed onto TNA (a), ENSO (b) and MSI (c) indices (in °C per standard deviation index). Panels (d) and (e) present anomalies of specific humidity at 850 hPa (colors, in kg/kg), DIV200/850 (contours, 0.3 10-6 s-1 CI) and moisture flux at 850 hPa (black and grey arrows for significant and non-significant anomalies respectively, kg/kg m/s) regressed onto TNA and ENSO indices respectively. Panel (f) corresponds to anomalies of specific humidity at 850 hPa (colors, in kg/kg), sea level pressure (contours, 3 Pa CI) and moisture flux at 850 hPa (black and grey arrows for significant and non-significant anomalies respectively, kg/kg m/s) regressed onto MSI index. Note that for panels (b) and (e), ENSO index is here multiplied by -1 in order to catch the la Niña effect. Contours in plots a and b delimit the 95% confidence interval using a Student t test, while in plots c and d only significant anomalies have been plotted. The continental box in (d), (e) and (f) delimits the Sahel region while boxes in panels (a), (b) and (c) correspond to TNA, ENSO and MSI indices areas respectively.

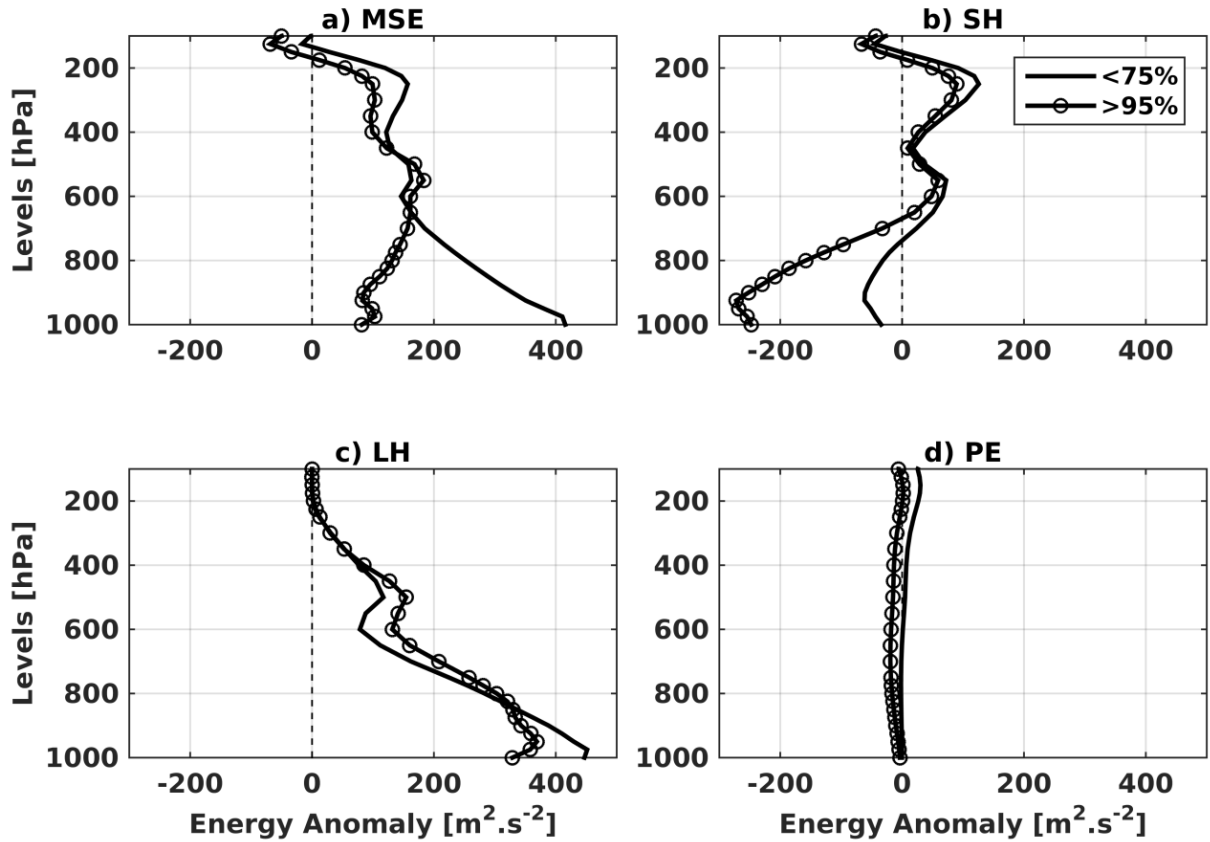

**Figure 5**: Regression of JJAS (1981-2016) anomalies of a) moist static energy (MSE) and its components: b) sensible heat (SH), c) latent heat (LH) and d) potential Energy (PE) onto moderate (line) and extreme (circles) normalized rainfall indices. The regressions onto the Sahel indices are done in each grid point of the Sahel box, and the vertical profiles correspond to anomalies averaged over the Sahel box.

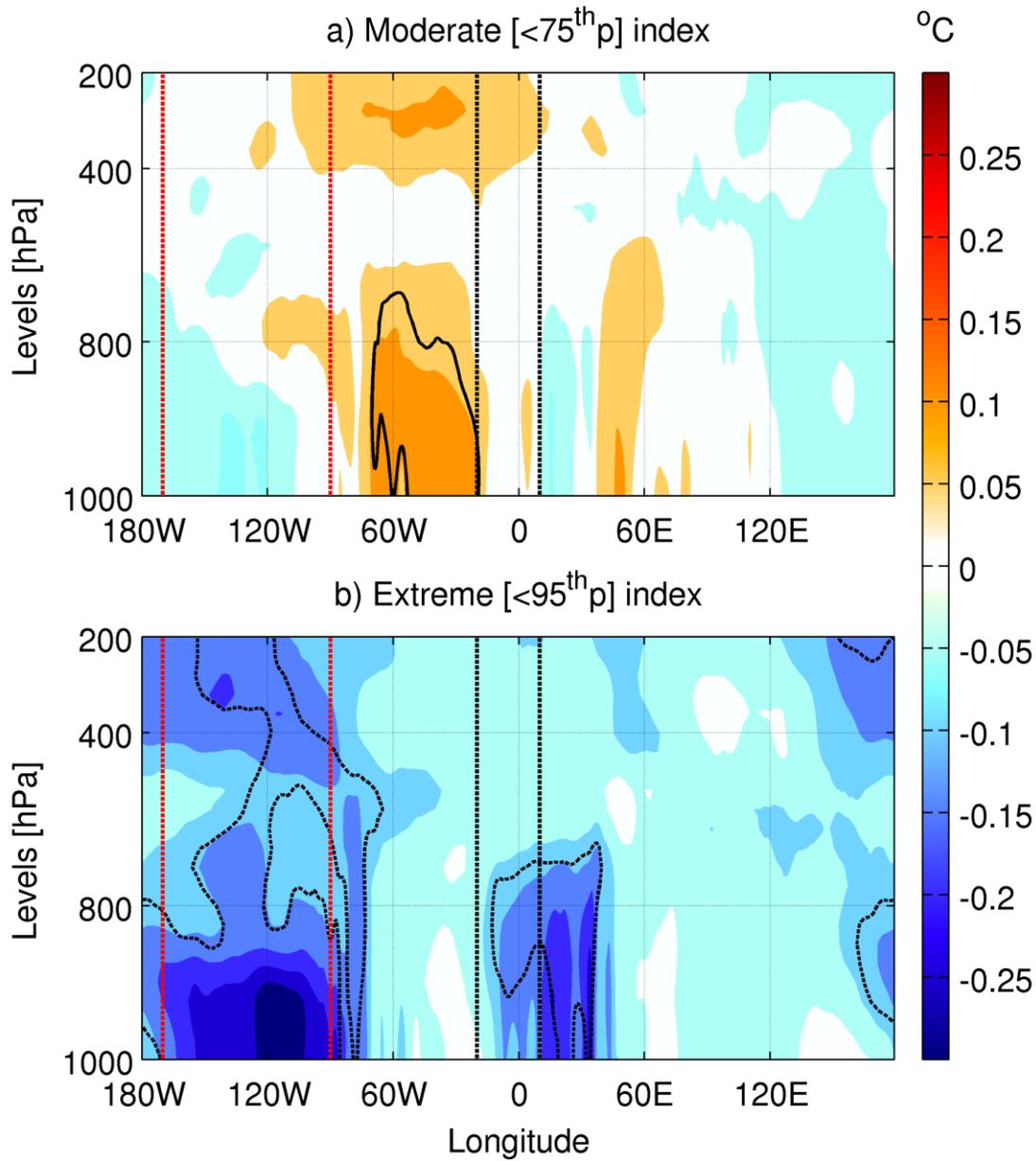

**Figure 6**: Regression of JJAS (1981-2016) vertical (1000-200 hPa) temperature anomalies onto moderate (a) and extreme (b) normalized rainfall indices. Temperature anomalies are averaged over the latitudinal corridor 0°-17°N for each pressure level (colors, in °C). Black dotted contours delimit significant anomalies at 95% confidence interval. Vertical black dotted bars delimit longitudinal boundaries of the Sahel region.

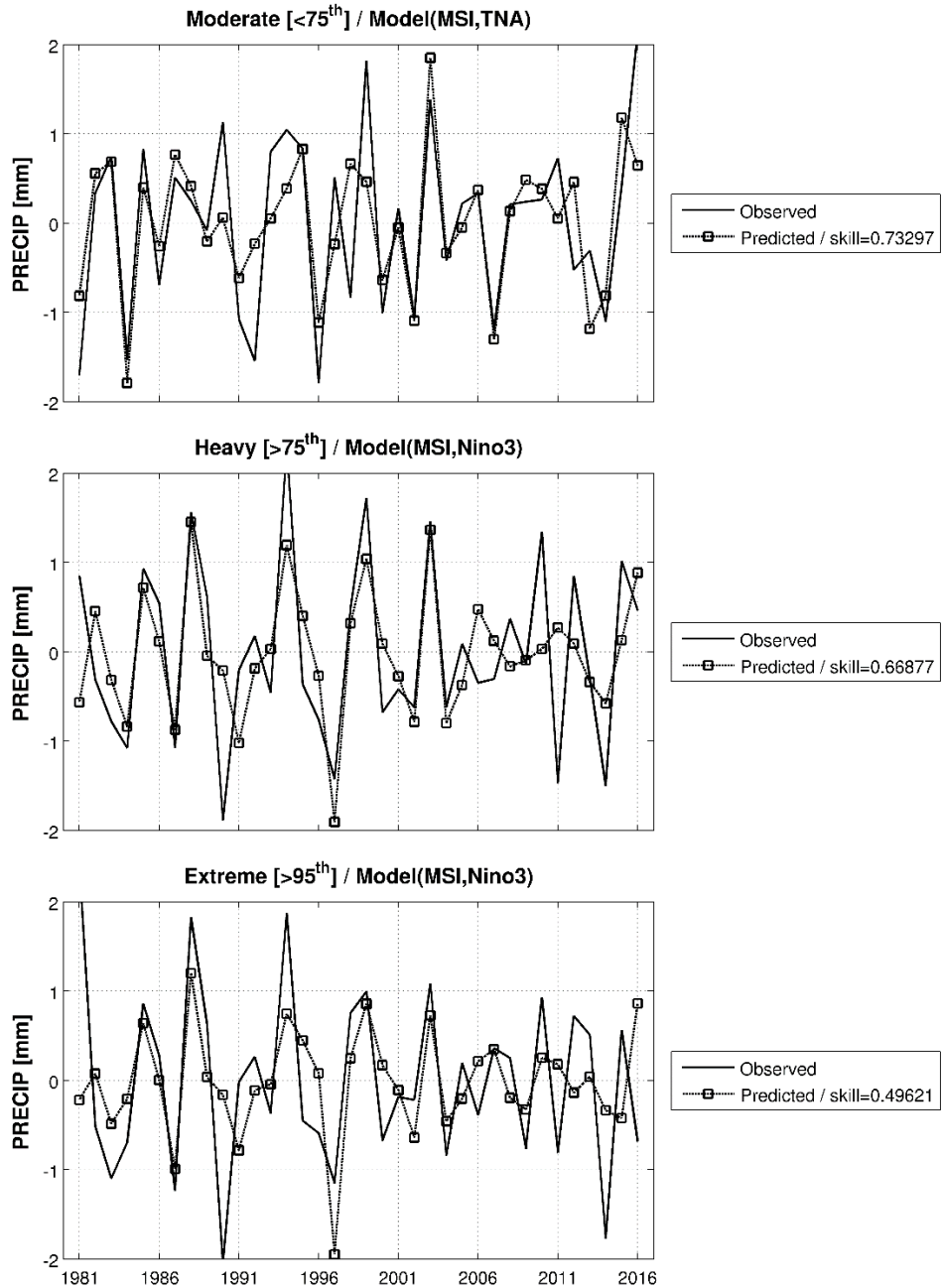

**Figure 7**: Cross-validated hindcast of Sahel August moderate (top), heavy (middle) and extreme (bottom) rainfall indices based on predictors selected from the stepwise regression (table 4). Note that the significant threshold for the model skill is equal to 0.2752.

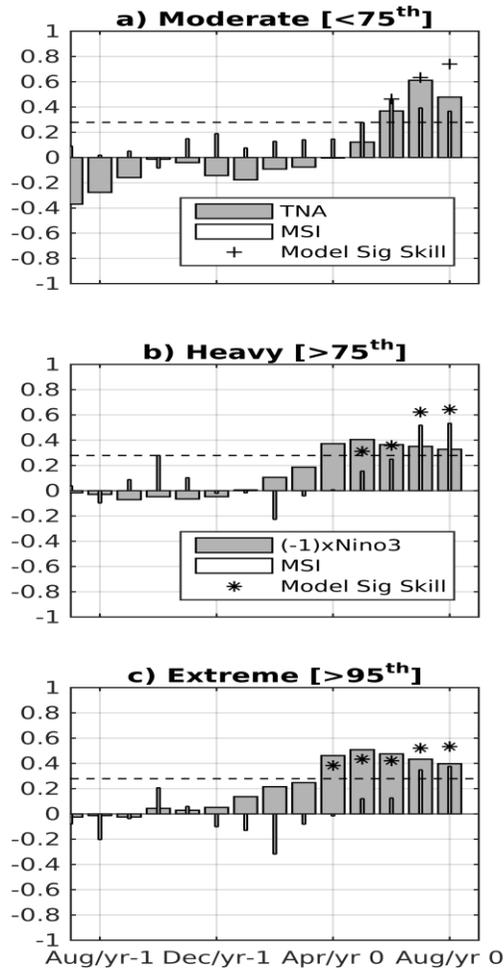

**Figure 8:** Lagged correlation and cross-validation hindcast of Sahel moderate (top), heavy (middle) and extreme (bottom) rainfall indices for August. Bars denote lagged correlations on a monthly basis between August precipitation indices (predictands) and preceding SST anomaly indices (predictors) from Table 4. The dashed horizontal line delimits the statistical significance (95% confidence level) for correlation. Markers are plotted whenever the regression model skill is statistically significant: crosses (predictors TNA and MSI considered), stars (predictors Nino3*(-1) and MSI considered). Note that the legend is the for panels (b) and (c).